\documentstyle[epsfig,aps,prl]{revtex}

\begin{document}
\twocolumn[\hsize\textwidth\columnwidth\hsize\csname
@twocolumnfalse\endcsname

\title{Echo spectroscopy and quantum stability of trapped atoms}
\author{M. F. Andersen, A. Kaplan and N. Davidson.}
\address{Department of Physics of Complex Systems, Weizmann Institute of Science,\\
Rehovot 76100, Israel }
\date{\today}
\maketitle

\begin{abstract}
We investigate the dephasing of ultra cold $^{85}Rb$ atoms trapped in an
optical dipole trap and prepared in a coherent superposition of their two
hyperfine ground states by interaction with a microwave pulse. We
demonstrate that the dephasing, measured as the Ramsey fringe contrast, can
be reversed by stimulating a coherence echo with a $\pi $-pulse between the
two $\frac{\pi }{2}$ pulses, in analogy to the photon echo. We also
demonstrate that the failure of the echo for certain trap parameters is due
to dynamics in the trap, and thereby that ''echo spectroscopy'' can be used
to study the quantum dynamics in the trap even when more than 10$^{6}$
states are thermally populated, and to study the crossover from quantum
(where dynamical decoherence is supressed) to classical dynamics.
\end{abstract}


]

The process of decoherence/dephasing is crucial to our understanding of the
connection between quantum mechanics and classical physics, since it is the
mechanism by which a pure quantum state evolves into a mixture of states
when the system is coupled to the environment. In the field of quantum
information long coherence times are of outmost importance, since
decoherence or dephasing represent loss of information, or decrease in
''fidelity''. Long coherence times of atomic ensembles prepared in
superposition states are of importance to high precision spectroscopy, where
the measurement time is a limiting factor on the precision of the
measurement. It is important to distinguish between reversible processes
(dephasing) and irreversible processes (decoherence). Dephasing can be
reversed, at least partially, by stimulating an effective ''time reversal'',
as has been reported for spin echoes \cite{Hahn50} and photon echoes \cite
{kurnit64}\cite{allen87}, and more recently for a motional wave packet echo
using ultra cold atoms in a one-dimensional optical lattice \cite
{Buchkremer00}.

Microwave (MW) spectroscopy has been previously performed on ultra cold
trapped atoms where long measurement times hold promise for high spectral
resolution\cite{davidson95}. However, dephasing due to inhomogeneous trap
pertubations limits the applicability of trapped atoms for precision
spectroscopy. Recently, a MW Ramsey technique was used to investigate the
coherence of accelerator modes in an atom-optical realization of the $\delta
$-kicked accelerator \cite{schlunk02}.

In this letter we investigate the dephasing of ultra cold $^{85}Rb$ atoms
trapped in an optical dipole trap and prepared in a coherent superposition
of the two hyperfine ground states by interaction with a \ so-called $\frac{%
\pi }{2}$-MW-pulse. We demonstrate that the dephasing, measured as the
Ramsey fringe contrast decay \cite{Ramsey56} and related with the Loschmidt
echo decay\cite{jalabert01}, can be reversed (for the proper trap
parameters), by stimulating a {\it coherence echo}. The later is obtained by
adding a population-inverting ''$\pi $'' pulse between the usual two ''$%
\frac{\pi }{2}$'' pulses. We also show that the failure of the echo\ for
other trap parameters is due to dynamics in the trap, and thereby that
''echo spectroscopy'' can be used to gain important information about the
time correlation function of the dynamics.

We study the two hyperfine levels of the ground state of $^{85}Rb$
atoms trapped in a dipole
trap. These two levels ($\left| 5S_{1/2},F=2,m_{F}=0\right\rangle$, denoted $%
\left| 1\right\rangle $, and $\left| 5S_{1/2},F=3,m_{F}=0\right\rangle $,
denoted $\left| 2\right\rangle $) are separated by the energy splitting $%
E_{HF}=\hbar\omega_{HF}$ with $\omega_{HF}=2\pi\times3.036$ GHz. Since the
dipole potential is inversely proportional to the trap laser detuning $%
\delta $ \cite{CCT91} there is a slightly different potential for atoms in
different hyperfine states \cite{note2}. This difference in potential is the
origin of the dephasing mechanisms we study here.

For spectroscopy of trapped atoms we must consider the entire
Hamiltonian including both the internal and the (center of mass)
motional degrees of freedom of the atom. The external potential
differs for the two levels and the Hamiltonian of our trapped two
level atom can be written as: \begin{figure}[b]
\epsfig{figure=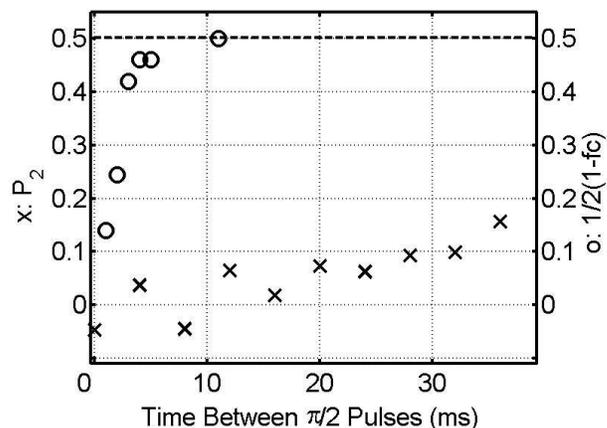, width=3.2in}
 \caption{1/2(1-fc)
where fc is the Ramsey fringe contrast (o) and echo signal $P_{2}$
measured as a function of time between $\protect\pi/2$-pulses (x).
The value 0 represents complete coherence, and the value 1/2
represents complete dephasing, for both Ramsey and echo signals.
Trap laser wavelength is 800 nm. A coherent echo ($P_{2}<<1/2$)
persists long after the Ramsey fringe contrast has decayed.}
\label{fi1}
\end{figure}%

\begin{figure}[tbp]
\mbox{\epsfig{figure=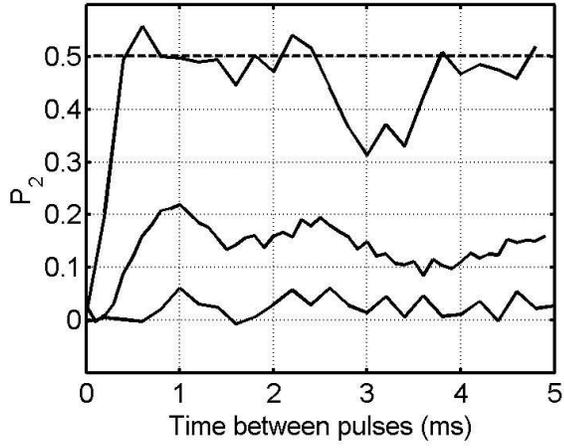, width=3in}} \caption{Echo signal
$P_{2}$ measured for three different trap laser wavelengths
$\protect\lambda $. Lower curve: $\protect\lambda $=805 nm. A good
echo signal ($P_{2}<<1/2$) is seen almost independent of time
between pulses. Middle curve: $\protect\lambda $=798.25 nm. The
echo signal
oscillates around a value smaller than 1/2 with partial revivals at $%
T=1/2T_{osc}$ and $T=T_{osc}$, where $T_{osc}$=3.6 ms is the
measured trap
oscillation frequency in the transverse direction. Upper curve: $\protect%
\lambda $=796.25 nm. After a short time the echo signal completely
disappears ($P_{2}=1/2$), but partly revives again at
$T=T_{osc}$.} \label{fi2}
\end{figure}

$H=H_{1}\left| 1\right\rangle \left\langle 1\right| +H_{2}\left|
2\right\rangle \left\langle 2\right| =\left(
\frac{p^{2}}{2m}+V_{1}\left( {\bf x}\right) \right) \left|
1\right\rangle \left\langle 1\right| +\left(
\frac{p^{2}}{2m}+V_{2}\left( {\bf x}\right) +E_{HF}\right) \left|
2\right\rangle \left\langle 2\right| $, where $p$ is the atomic
center of mass momentum and $V_{1}$ [$V_{2}$] the external
potential for an atom in state $\left| 1\right\rangle $ [$\left|
2\right\rangle $], much smaller than $E_{HF}$. The atoms are initially prepared in their internal ground state $%
\left| 1\right\rangle $. Their total wavefunction can be written as $\Psi
=\left| 1\right\rangle \otimes $ $\psi $, where $\psi $ represents the
motional (external degree of freedom) part of their wavefunction. If a MW
field close to resonance with $\omega _{HF}$ is applied, transitions between
the eigenstates of the Hamiltonian corresponding to different internal
states, can be driven. Since the size of our trap ($\sim 50$ $\mu m$) is
much smaller than the MW wavelength ($\sim $10 cm), the momentum of the MW
photon can be neglected (Lamb-Dicke regime \cite{CCT91}). The transition
matrix elements are given by $C_{nn^{\prime }}=\left\langle n^{\prime }\mid
n\right\rangle \times M_{1\rightarrow 2}$ where $M_{1\rightarrow 2}$ is the
free space matrix element for the internal state transition, and $%
\left\langle n^{\prime }\mid n\right\rangle $ is the overlap between the
initial motional eigenstate of $H_{1}$ and an eigenstate of $H_{2}$. When a
strong and short MW pulse is applied, the motional part of the initial
wavefunction is simply projected into $V_{2}\left( {\bf x}\right) $. If $%
V_{1}\left( {\bf x}\right) =V_{2}\left( {\bf x}\right) $ then clearly $%
\left\langle n^{\prime }\mid n\right\rangle =\delta _{nn^{\prime }}$. For a
small enough ''perturbation'', i.e. $0<\delta V=V_{2}\left( {\bf x}\right)
-V_{1}\left( {\bf x}\right) $ $<$ $\varepsilon $, for some $\varepsilon $ to
be determined, we will still have $\left\langle n\mid n^{\prime
}\right\rangle \simeq \delta _{nn^{\prime }}$. In the general case, a
projected eigenstate of $H_{1}$ will not be an eigenstate of $H_{2}$, and
therefore will evolve in the new potential, causing the overlap $\left|
\left\langle n\left( t=0\right) \mid n\left( t=\tau \right) \right\rangle
\right| $ to decay ($\left| n\left( t=\tau \right) \right\rangle \equiv \exp
\left( -i\frac{H_{2}}{\hbar }\tau \right) $ $\left| n\right\rangle $).

In Ramsey spectroscopy, two ''$\pi /2$'' pulses are applied, separated by a
variable time $\tau $. If we start with some eigenstate of $V_{1}$
characterized by the quantum number $n$, then the probability to be in the
internal state $\left| 2\right\rangle $ after the $\pi /2$-$\pi /2$ pulse
sequence can be shown to be: $P_{2}=\frac{1}{2}\left( 1+\left\langle n\left|
e^{-i\left( \frac{H_{2}}{\hbar }-\omega _{MW}+\Delta _{n}\right) \tau
}\right| n\right\rangle \cos \left( \Delta _{n}\tau \right) \right) $ where $%
\omega _{MW}$ is the MW-frequency, $\tau $ is the time between the pulses,
and $\Delta _{n}$ is a generalized, state-dependent, detuning (it reduces to
the detuning when only two levels are coupled by the MW-field) defined so $%
\left\langle n\left| \exp \left( -i\left( \frac{H_{2}}{\hbar }-\omega
_{MW}+\Delta _{n}\right) \tau \right) \right| n\right\rangle $ is real and
positive. Scanning $\omega _{MW}$ for a fixed $\tau $ yields the usual
Ramsey fringes with a contrast given by $\left\langle n\left| \exp \left(
-i\left( \frac{H_{2}}{\hbar }-\omega _{MW}+\Delta _{n}\right) \tau \right)
\right| n\right\rangle =\left| \left\langle n\left( t=0\right) \mid n\left(
t=\tau \right) \right\rangle \right| $. The fringe contrast can be viewed as
the survival probability for a state $\left| n\right\rangle $ after a sudden
change in potential from $V_{1}\left( {\bf x}\right) $ to $V_{2}\left( {\bf x%
}\right) $. Also, since $\left\langle n\left| \exp \left( -i\left( \frac{%
H_{2}}{\hbar }-\omega _{MW}+\Delta _{n}\right) \tau \right) \right|
n\right\rangle =\left| \left\langle n\left| \exp \left( i\left( \frac{H_{1}}{%
\hbar }\right) \tau \right) \exp \left( -i\left( \frac{H_{2}}{\hbar }\right)
\tau \right) \right| n\right\rangle \right| $ it can be related to the
Loshmidt echo, as described in Ref. \cite{jalabert01}, and serves as a
measure of stability \cite{peres84}.

In our experiment we don't have an initial single motional eigenstate, but a
thermal ensemble of atoms incoherently populating more than $10^{6}$
eigenstates. The total population in $\left| 2\right\rangle $ will be given
by an average of $P_{2}$ over the initial thermal ensemble. Since $\Delta
_{n}$ depends on the initial state, the fringe contrast of the
ensemble-averaged $P_{2}$ decays rapidly even when $\left| \left\langle
n\left( t=0\right) \mid n\left( t=\tau \right) \right\rangle \right| \simeq
1 $ for all populated states.

Our experiment is as follows: $^{85}Rb$ atoms are loaded into a far off
resonance optical trap (FORT), cooled to a temperature of 20 $\mu K$, and
optically pumped into the F=2 hyperfine state from a magneto optical trap,
followed by a molasses stage. The FORT consists of a 50 mW horizontal laser
beam focused to a 1/e$^{2}$ radius of 50 $\mu m$, and with a wavelength of $%
\lambda $=800 $nm$ ($\sim 5$ $nm$ from the $D_{1}$ line) yielding a trap
depth of $U_{0}/k_{b}T=1.5$. The longitudinal oscillation time in the FORT ($%
\sim 500$ ms) is much larger than the experiment time (of the order of ten
ms), and hence only the transverse motion is considered. The transverse
oscillation time was measured using parametric excitation spectroscopy \cite
{friebel98} to be 3.6 ms, in good agreement with the calculated value of 3
ms. For these trap parameters $\left\langle n^{\prime }\mid n\right\rangle
\simeq \delta _{n,n^{\prime }}$, and therefore $\left| \left\langle n\left(
t=0\right) \mid n\left( t=\tau \right) \right\rangle \right| \simeq 1$ for
all thermally populated states. The free space Rabi-frequency of the MW
fields is 5 kHz, so the duration of the MW-pulses can be neglected. A bias
magnetic field of 40 mG shifts all m$_{F}\neq 0$ states out of resonance
with the MW field, limiting the MW transitions to the two m$_{F}=0$ states ($%
\left| 1\right\rangle $ and $\left| 2\right\rangle $). After the MW pulses
the population of state $\left| 2\right\rangle $ is detected by normalized
selective fluorescence detection \cite{Khaykovich00}, and the Ramsey fringe
contrast is normalized to the number of atoms in $\left| 2\right\rangle $
with a short $\pi $-pulse.

As seen in Fig. \ref{fi1}, the Ramsey fringe contrast decays on a
time scale of a few ms, due to the variation of $\Delta _{n}$ over
the thermally populated states in good
agreement\begin{figure}[tbp] \mbox{\epsfig{figure=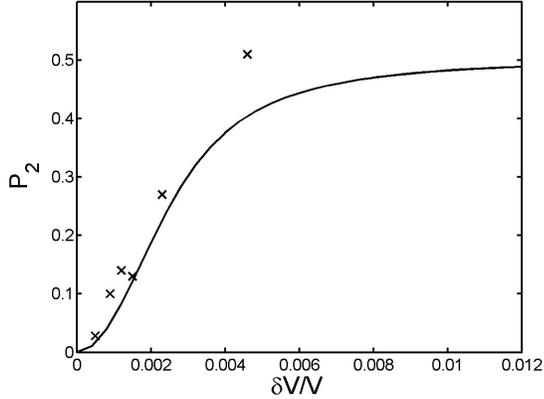,
width=3in}} \caption{x: long time level of measured echo signal as
a function of difference between potentials ($\protect\delta
V/V$). Solid line: Calculation of the ensemble average of $\left|
\left\langle n^{\prime}=n \mid n\right\rangle \right| ^{4}$ (a
measure of the quantum stability of the system) for a 2D harmonic
potential with oscillation time of 3.6 ms \protect\cite{note3}.}
\label{fi3}
\end{figure}with a calculated
decay time of 2.7 ms. The decay time is calculated as 1/(2$\Delta
_{RMS}$), where $\Delta _{RMS}$ is the RMS of the resonance
frequency for $\left| n\right\rangle \rightarrow \left| n^{\prime
}=n\right\rangle $ transitions, taken over a thermal ensemble in a
harmonic trap clipped at 1.5 $k_{B}T$. As explained above, the
observed decay of the Ramsey fringe contrast is not a fingerprint
of decoherence, but a consequence of dephasing. This dephasing can
be reversed by adding a MW $\pi $-pulse between the two $\frac{\pi
}{2}$ pulses, which inverts the populations of $\left|
1\right\rangle $ and $\left| 2\right\rangle $. $P_{2}$ then
becomes:
\begin{eqnarray}
P_{2} &=&\frac{1}{2}\left[ 1-%
\mathop{\rm Re}%
\left( \left\langle n\left| e^{i\left( \frac{H_{1}}{\hbar }\right) \tau
}e^{i\left( \frac{H_{2}}{\hbar }\right) \tau }e^{-i\left( \frac{H_{1}}{\hbar
}\right) \tau }e^{-i\left( \frac{H_{2}}{\hbar }\right) \tau }\right|
n\right\rangle \right) \right]  \nonumber \\
&=&\frac{1}{2}\left[ 1-%
\mathop{\rm Re}%
\left( e^{i\frac{E_{n}^{1}}{\hbar }\tau }\left\langle \varphi _{n}\left(
t=0\right) \mid \varphi _{n}\left( t=\tau \right) \right\rangle \right) %
\right]  \label{formel1}
\end{eqnarray}
where $\left\{ \varphi _{n}\right\} $ is a new basis defined by $\left|
\varphi _{n}\left( t=0\right) \right\rangle \equiv \exp \left( -i\frac{H_{2}%
}{\hbar }\tau \right) $ $\left| n\right\rangle $, and $\left| \varphi
_{n}\left( t=\tau \right) \right\rangle \equiv \exp \left( -i\frac{H_{1}}{%
\hbar }\tau \right) $ $\left| \varphi _{n}\left( t=0\right) \right\rangle $%
\cite{note1}. $P_{2}$ no longer depends on $\omega _{MW}$ and $E_{HF}$. The
condition $\left\langle n^{\prime }\mid n\right\rangle \simeq \delta
_{nn^{\prime }}$ for all initially populated vibrational states now ensures
that $P_{2}\simeq 0$ for all $\tau $. In other words, after dephasing for a
time $\tau $ a $\pi $-pulse stimulates a {\it coherence echo }at time 2$\tau
$. The results of $\pi /2$-$\pi $-$\pi /2$ echo spectroscopy with the same
trap parameters as before are also presented in Fig \ref{fi1}. A coherence
echo ($P_{2}<<1/2$) is clearly seen long after the Ramsey fringe contrast
has decayed. On a time scale of $\sim $100 ms the echo coherence decays,
presumably due to longitudinal motion.

When $\left\langle n^{\prime }\mid n\right\rangle \neq \delta _{nn^{\prime
}} $ a good echo signal is no longer expected, since each vibrational state
is coupled to several vibrational states by the MW fields, and therefore $%
\left| \left\langle \varphi _{n}\left( t=0\right) \mid \varphi _{n}\left(
t=\tau \right) \right\rangle \right| <1$. We perform echo spectroscopy as a
function of time between pulses for different wavelengths of the trap laser,
thereby changing the perturbation strength $\delta V$ \cite{note2}, while
keeping the trap depth constant by adjusting the power of the trap beam. The
results are shown in Fig. \ref{fi2}. For a large detuning ($\lambda $=805
nm, 10 nm from the $D_{1}$ line at 795 nm) a good echo ($P_{2}<<1/2$) is
seen independent of the time between pulses, as also seen in Fig. \ref{fi1}.
For an intermediate detuning ($\lambda $=798.25 nm) damped oscillations to a
level smaller than 1/2 are seen, and for a small detuning ($\lambda $=796.25
nm) a complete decay of the echo coherence ($P_{2}=1/2$) followed by partial
revivals at later times is seen.

The interpretation of the large detuning regime was given above. For
intermediate detunings a small but significant coupling to other ($n^{\prime
}\neq n$) vibrational states exists. Ignoring for a moment gravity, $%
V_{1}\left( {\bf x}\right) $ and $V_{2}\left( {\bf x}\right) $ are just the
optical potentials and are related by $V_{2}^{{}}\left( {\bf x}\right)
=\left( 1+\varepsilon \right) V_{1}^{{}}\left( {\bf x}\right) $ \cite{note2}%
. When atoms are transferred from $V_{1}\left( {\bf x}\right) $ to $%
V_{2}\left( {\bf x}\right) $ by the MW field, a parametric excitation of the
atomic wavepacked is induced, thereby exiting breathing modes. However, the
change in the optical potential also changes the gravitational sag, and
hence excites sloshing modes along the vertical axis. A wavefunction
parametrically exited in an harmonic oscillator will revive after $\tau
=m\times 1/2\tau _{osc}$ (yielding $\left| \left\langle \varphi _{n}\left(
t=0\right) \mid \varphi _{n}\left( t=m\times 1/2\tau _{osc}\right)
\right\rangle \right| =1$), and a sloshing mode will revive after $\tau
=m\times \tau _{osc}$ \cite{Buchkremer00}\cite{ejnisman97}\cite{Raithel98}.
These revivals are seen in Fig. \ref{fi2} as a partial revival in the echo
signal at $\tau =1.8$ ms and a better one at $\tau =3.6$ ms, and their
origin was verified in numerical calculations of the echo signal for typical
single states in a 2D harmonic oscillator. When gravity was omitted in the
calculation, the revival at $\tau =1.8$ ms was complete. The damping of the
oscillation of the echo signal, i.e. the lack of a perfect revival at $\tau
=\tau _{osc}$, is due to the anharmonicity of the Gaussian trap. We stress
the sensitivity of our technique to map the quantum dynamics of a system,
since we see the dynamics due to a pertubation (kick), that is about three
orders of magnitude smaller than $k_{B}T$. We note the similarity of this
technique, to techniques used to map molecular potentials in chemistry as
e.g. COIN \cite{leichtle98}.

For sufficiently long time $\tau $ the wave packet oscillations of Fig. \ref
{fi2} damp due to complete dephasing of the dynamics. At such long time a
simple expression for the echo signal can be given. In particular assuming
random phases between all vibrational states yields the simple relation $%
\left| \left\langle \varphi _{n}\left( t=0\right) \mid \varphi _{n}\left(
t=\tau \right) \right\rangle \right| =\left| \left\langle n^{\prime }=n\mid
n\right\rangle \right| ^{4}$. Substituting this into Eq. \ref{formel1} and
averaging over the ensemble yield the expected long-time echo signal. We
performed this calculation numerically for a 2D harmonic trap, in gravity,
with our measured oscillation frequency and a thermal ensemble with a
temperature of 20 $\mu K$ clipped at our trap depth of 1.5$k_{B}T$ \cite
{note3}. The results are shown in Fig. \ref{fi3}, together with the measured
long-time echo signals as a function of potential difference. As seen, the
calculation for the harmonic trap and the data points for the Gaussian trap
show the same qualitative behavior, of improved long time echo when $\delta
V $ becomes small.

\begin{figure}[tbp]
\mbox{\epsfig{figure=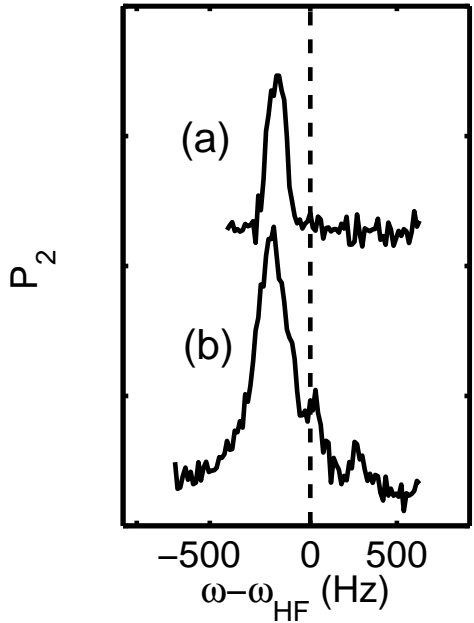, width=2.5in}}
\caption{MW-spectroscopy of atoms in a trap at a wavelength of 805
nm, with
a 20 ms MW-pulse. MW-power equivalent of (a): a free space $\protect\pi $%
-pulse, (b): a free space 4$\protect\pi $-pulse.} \label{fi4}
\end{figure}

Finally, non vanishing matrix elements for transitions to vibrational states
with $n^{\prime}\neq n$ should show up as sidebands in the MW spectrum, if
the pulse is long and weak enough so that the power broadening is smaller
than the typical level spacing. This is shown in Fig. \ref{fi4} for a trap
laser wavelength of 805 nm. For a MW power corresponding to a free space $%
\pi $-pulse (Fig. \ref{fi4}a) the sidebands are not visible, indicating that
the matrix elements for the sidebands are very small and thus enabling a
successful echo in agreement with Fig. \ref{fi2}. Only for the much stronger
4$\pi$-pulse (Fig. \ref{fi4}b), the sidebands emerge. We also verified that
the sidebands are stronger for smaller detunings, as expected from the
analysis. This is interesting because it allows the quantization of the trap
levels to be detected, even at very large temperatures compared to the 1D
level spacing ($k_{B}T>10^{3}\ast\hbar\omega_{osc}$), where the motion of
particles is usually considered to be classical.

In summary, we have demonstrated that a macroscopic coherence, lost as a
consequence of dephasing of the different populated motional levels, can be
efficiently revived by stimulating an echo if the trap detunings is large
enough so $\left| \left\langle n^{\prime }=n\mid n\right\rangle \right|
\simeq 1$. This suppression of dephasing due to pertubations induced by the
trap yields a dramatic increase in the coherence time for trapped atoms,
that may find important applications for precision spectroscopy and quantum
information processing. We also demonstrated that echo spectroscopy can be
used to experimentally map the quantum dynamics of trapped atoms, namely the
ensemble average of $\left| \left\langle \varphi _{n}\left( t=0\right) \mid
\varphi _{n}\left( t=\tau \right) \right\rangle \right| $, showing a
crossover between quantum and classical regimes. In this way it can serve as
an extremely sensitive experimental tool for investigating quantum chaos in
atom-optics billiards \cite{milner01} even when more than 10$^{6}$ states
are thermally populated. Finally, we demonstrated that for sufficiently
large detunings the sidebands can be resolved, thus making it possible to
select mono energetic atoms by help of MW pulses, and thereby enabling
energy-dependent repumping with MW pulses, for new cooling schemes.


\begin{references}
\bibitem{Hahn50}  E. L. Hahn, Phys. Rev. {\bf 80}, 580 (1950).

\bibitem{kurnit64}  N. A. Kurnit, I. D. Adella and S. R. Hartmann, Phys Rev.
Lett. {\bf 13 }567 (1964).

\bibitem{allen87}  L. Allen and J. H. Eberly, {\it Optical Resonance and
Two-Level Atoms,} Dover Publications, Inc., New York, (1987).

\bibitem{Buchkremer00}  F. B. J. Buchkremer, R. Dumke, H. Levsen, G. Birkl
and W. Ertmer, Phys. Rev. Lett. {\bf 85} 3121 (2000).

\bibitem{davidson95}  N. Davidson, H. J. Lee, C. S. Adams, M. Kasevich and
S. Chu, Phys. Rev. Lett. {\bf 74,} 1311 (1995).

\bibitem{schlunk02}  S. Schlunk et. al., arXiv:physics/0207075 (2002).

\bibitem{Ramsey56}  N. F. Ramsey, {\it Molecular Beams}, Oxford at the
Clarendon Press (1956).

\bibitem{jalabert01}  R. A. Jalabert and H. M. Pastawski, Phys. Rev. Lett.,
{\bf 86}, 2490 (2001).

\bibitem{CCT91}  C. Cohen-Tannoudji, J. Dupont-Roc and G. Grynberg, {\it %
Atom-Photon interactions,} John Wiley \& Sons Inc. (1992).

\bibitem{note2}  For these two hyperfine levels the matrix element for the
dipole interaction are identical, hence the relative difference between the
two optical potentials is $\left( V_{1}-V_{2}\right) /V_{2}\simeq\frac{%
\omega_{HF}}{\delta}\sim10^{-3}$ for our experimental parameters.

\bibitem{peres84}  A. Peres, Phys. Rev. A {\bf 30}, 1610 (1984).

\bibitem{friebel98}  S. Friebel, C. DAndrea, J. Walz, M. Weitz, and T. W.
Hansch, Phys. Rev. A {\bf 57}, R20 (1998).

\bibitem{Khaykovich00}  L. Khaykovich, N. Friedman, S. Baluschev, D. Fathi
and N. Davidson, Europhys. Lett., {\bf 50}, 454 (2000).

\bibitem{ejnisman97}  R. Ejnisman, P. Rudy, H. Pu and N. P. Bigelow, Phys.
Rev. A., {\bf 56,} 4331 (1997).

\bibitem{Raithel98}  G. Raithel, W. D. Phillips, S. L. Rolston, Phys. Rev.
Lett. {\bf 81, }3615 (1998).

\bibitem{leichtle98}  C. Leichtle, W. P. Schleich, I. Sh. Averbukh and M.
Shapiro, J. Chem. Phys. {\bf 108}, 6057 (1998).

\bibitem{note3}  The ensemble average for the clipped thermal distribution
in the harmonic trap was calculated as: $\sum\limits_{n,m}\left(
\left| \left\langle u_{m}^{v2}\mid u_{m}^{v1}\right\rangle \right|
\left| \left\langle u_{n}^{h2}\mid u_{n}^{h1}\right\rangle \right|
\right) ^{4} e^{-E_{nm}/k_{B}T} /\sum\limits_{n,m}
e^{-E_{nm}/k_{B}T} $ where $u_{n}^{h1}$ are eigenfunction of the
harmonic
oscillator associated with the horizontal motion and $\left| 1\right\rangle $%
, $u_{n}^{h2}$ are associated with $\left| 2\right\rangle $ (the spring
constant is slightly stronger). $u_{m}^{v2}$ and $u_{m}^{v1}$ are associated
with the vertical motion where there is both a change in spring constant and
a change in zero point due to different gravitational sag. The sum is over
the $\sim$3$\times10^{6}$ states for which $E_{nm}=\left( n+m\right) \hbar
\omega_{osc}<U_{pot}$.

\bibitem{milner01}  V. Milner, J. L. Hanssen, W. C. Campbell and M. G.
Raizen, Phys. Rev. Lett. {\bf 86}, 1514 (2001) and N. Friedman, A. Kaplan,
D. Carasso and N. Davidson, Phys. Rev. Lett. {\bf 86}, 1518 (2001)

\bibitem{note1}  $%
\mathop{\rm Re}%
\left( \exp\left( i\frac{E_{n}^{1}}{\hbar }\tau\right) \left\langle
\varphi_{n}\left( t=0\right) \mid\varphi _{n}\left( t=\tau\right)
\right\rangle \right) \simeq\left| \left\langle \varphi_{n}\left( t=0\right)
\mid\varphi_{n}\left( t=\tau\right) \right\rangle \right| $ is a good
approximation for small enough perturbation, as was confirmed in numerical
calculations, for a 2D harmonic oscillator.
\end{references}
\end{document}